\documentstyle[12pt,epsfig]{article}

\newcommand{\be}{\begin{equation}}
\newcommand{\ee}{\end{equation}}

 1
\font\elevenrm=cmr10 scaled\magstep 1
 1

\def\reff{\hang\noindent}

\begin{document}
\vspace*{1.8cm}
  \centerline{\bf MULTIWAVELENGTH OBSERVATIONS OF CLUSTERS OF GALAXIES AND}
  \centerline{\bf THE ROLE OF CLUSTER MERGERS}
\vspace{1cm}
  \centerline{PASQUALE BLASI}
\vspace{1.4cm}
  \centerline{NASA/Fermilab Astrophysics Center}
  \centerline{\elevenrm Fermi National Accelerator Laboratory, Box 500,
Batavia, IL 60510-0500, USA}
\vspace{3cm}
\begin{abstract}
Some clusters of galaxies have been identified as powerful sources
of non-thermal radiation, from the radio to X-ray wavelengths. 
The classical models proposed for the explanation of this radiation 
usually require large energy densities in cosmic rays in the intracluster
medium and magnetic fields much lower that those measured using the
Faraday rotation. We study here the role that mergers of clusters of
galaxies may play in the generation of the non-thermal radiation,
and we seek for additional observable consequences of the model. 
We find that if hard X-rays and radio radiation are respectively interpreted as
inverse Compton scattering (ICS) and synchrotron emission of relativistic
electrons, large gamma ray fluxes are produced, and for the Coma cluster,
where upper limits are available, these limits are exceeded. 
We also discuss an alternative and testable model that naturally solves
the problems mentioned above.

\end{abstract}

\vspace{2.0cm}

\section{Introduction}

Non-thermal radiation is observed in several clusters of galaxies at 
frequencies varying from the radio [see (Ensslin 1999) for a recent review]
to the UV to the soft and hard 
X-rays. These observations are hard to reconcile with our classical view
that would interpret the radio radiation as synchrotron emission and
the hard X-rays and UV radiation as ICS of the same electrons off the
photons of the cosmic microwave background. This 
interpretation in general requires a large rate of injection of cosmic
rays at recent times and intracluster magnetic fields smaller than 
those obtained by Faraday rotation measurements (Eilek 1999; Clarke et
al. 1999) by a factor $\sim 50-100$. 
These can be considered at least circumstantial evidence for some problems
in the {\it conventional interpretation}, although additional observational 
tests are required in order to have more solid proofs. 

The only events that seem to be able to provide the required energetics are 
mergers of clusters of galaxies. In this paper we study in detail all the
channels that may contribute to the production of non-thermal radiation 
during a cluster merger, including the contribution of both primary and
secondary electrons. Since in some of the clusters observed in radio
and X-rays there does not seem to be evidence for an ongoing merger, we 
concentrate our attention on mergers occurred in the past. 
Although the calculations are carried out in the most general form, 
we apply them specifically to the case of the Coma cluster, for which 
a complete set of multiwavelength observations have been carried out.
A more general discussion of our results and more technical details can
be found in (Blasi 2000). 

For the case of Coma, fitting the non-thermal multiwavelength observations
has the following implications: 1) Magnetic fields $\sim 50$ times
smaller than the ones measured by Eilek (1999) and Clarke et al. (1999)
are necessary; 2) the gamma ray flux at $\sim 100$ MeV exceeds the EGRET 
upper limits; 3) the merger must have been just ended.

Since the main reason for these problems resides in the 
synchrotron plus ICS model, some attempts have been made to look for
alternative interpretations of the hard X-ray excess, in particular invoking
the bremsstrahlung emission from a non-thermal tail in the electron
distribution, produced by stochastic acceleration (Blasi 2000a; Ensslin,
Lieu and Biermann 1999; Dogiel 1999).
The X-ray spectra can be fitted by this model without
requiring small magnetic fields, although large injection rates of MHD waves 
are needed (compatible with the expectations in a merger of two
clusters of galaxies).

The paper is structured as follows: in section 2 we describe mergers
as particle accelerators; in section 3 we describe our calculations;
in section 4 we discuss an alternative model of the non-thermal radiation
produced during cluster mergers and its consequences. We conclude in section 5.

\section{Cluster mergers as particle accelerators}

During the merger between two clusters of galaxies, a typical gravitational
energy of $E_{merger}\sim 1.4\times 10^{64}$ erg is available (we assumed here
a typical total mass of a cluster to be $\sim 5\times 10^{14} M_\odot$ and
a distance of $\sim 1.5$ Mpc between the two clusters). During the approach,
this energy is mainly converted into kinetic energy of the dark matter 
component, which is weakly interacting. However, at some point a strong shock
is formed in the baryonic component and energy is transferred from dark matter
to baryons (and electrons). This process is thought to produce the heating
of the intracluster gas. A small fraction of this energy however can be 
converted into kinetic energy of particles out of equilibrium (non-thermal)
through first order Fermi acceleration at the shock. Both electrons and
protons are expected to be accelerated although the process should be more
efficient for protons. A detailed description of this problem can be found in
(Levinson 1994; McClements et al. 1997), but a brief description can be 
useful here: protons with low energy do
interact with Alfv\'en waves, while electrons do not. Therefore protons can 
be more easily injected into the accelerator while electrons need to interact
with some other kind of waves (for instance whistlers). Moreover, in order
for a particle to be accelerated at the shock, its Larmor radius must be 
larger than the width of the shock, comparable with the Larmor radius 
of thermal protons. It is easy to see that this condition is realized only for
electrons with energy larger than $5-10$ MeV (much larger than the typical
cluster temperature). Electrons need therefore a preacceleration, otherwise 
the fraction of electrons injected in the shock region remains insignificantly
small. We introduce here the coefficient $\xi<1$ as the ratio of the spectra
of electrons and protons at injection at fixed energy. For the injection
spectrum of electrons and protons we use here a power law in momentum with
index $\gamma=2.32$, needed to fit the radio spectrum of Coma (this value is
compatible with the compression ratios at the merger's shocks observed in
the simulations of Takizawa and Naito (2000)). These spectra are changed 
by propagation and losses effects, 
which we calculate by solving the full transport equations for primary 
electrons, protons and secondary electrons, generated by the decay of the
charged pions due to $pp$ inelastic scattering in the intracluster medium
[see (Sarazin 1999) for a detailed description of the energy losses and
radiative properties of primary electrons].

The cosmic ray confinement of the proton component over cosmological scales 
(Berezinsky, Blasi and Ptuskin 1997;Volk, Aharonian and Breitschwerdt 1996;
Colafrancesco and Blasi 1998) and the fast energy losses of electrons 
make the results very weakly dependent
on the choice of the diffusion coefficient, with the exception of the maximum 
energies achievable in the acceleration. We use a diffusion coefficient derived
by applying the quasi linear theory to a Kolmogorov spectrum of magnetic
fluctuations (Blasi and Colafrancesco 1999) with the size of the largest 
eddy $\sim 200$ kpc (typical size
between two galaxies). Higher maximum energies can be attained adopting a 
Bohm diffusion coefficient, but the fluxes far away from the tails do not
change appreciably, as stressed above. 

\section{The calculations}

Our calculations are fully time-dependent, therefore we can evaluate
the non-thermal spectra at each time during or after the merger. Some 
general findings are the following: {\it i)} sufficiently high energy 
electrons, required to generate radio radiation and hard X-rays, exist 
in the cluster only if the merger has just ended or is currently ongoing. 
In particular the higher 
frequencies in the radio spectrum can be produced only if the merger was 
not over more than $\sim 20$ million years ago. 
{\it ii)} The secondary electron component is time independent: in fact the
protons injected during the merger are confined in the intracluster medium
and serve as a continuous source of new electrons, even if the proton injection
ended long ago. {\it iii)} The gamma
ray fluxes due to secondary electrons or due to pion decays are time
independent. They cannot be washed out by the time evolution.

\begin{figure}[thb]
 \begin{center}
  \mbox{\epsfig{file=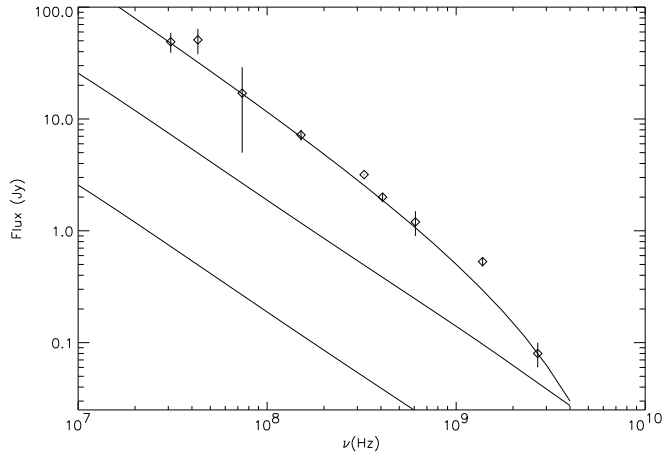,width=9.cm}}
  \caption{\em {Spectrum of radio radiation from Coma. 
}}
 \end{center}
\end{figure}

On this basis, we assume here that the merger just ended in Coma and
calculate the fluxes of non-thermal radiation.
The results for the radio emission of Coma are plotted in Fig. 1, where 
a magnetic field of $\sim 0.15\mu G$ was required. The thick curve that 
fits the observations (Feretti et al. 1995) refers to synchrotron emission 
of primary electrons. The cutoff is due to the maximum energy of the 
electrons and 
therefore might be in a different place for different choices of the 
diffusion coefficient. The real presence of a cutoff at $>2$ GHz was questioned
anyway by Deiss et al. (1997).
The two thin lines represent the synchrotron contribution of the secondary 
electrons for $\xi=0.1$ (lower curve) and $\xi=0.01$ (upper curve). 

The results of our calculations for the hard X-rays and UV-soft X-ray 
radiation are plotted in Fig. 2, together with the expectation due to
thermal bremsstrahlung of a gas of electrons at the temperature of Coma 
(thick solid curve). The data points are from BeppoSAX (Fusco-Femiano et al. 
1999). The thick band is an estimate of the UV excess (Lieu et al. 1999). 
The thin solid line is the
ICS contribution of primary electrons while the two dashed lines
represent the ICS radiation of secondary electrons for $\xi=0.1$ (lower
curve) and $\xi=0.01$ (upper curve). 

\begin{figure}[thb]
 \begin{center}
  \mbox{\epsfig{file=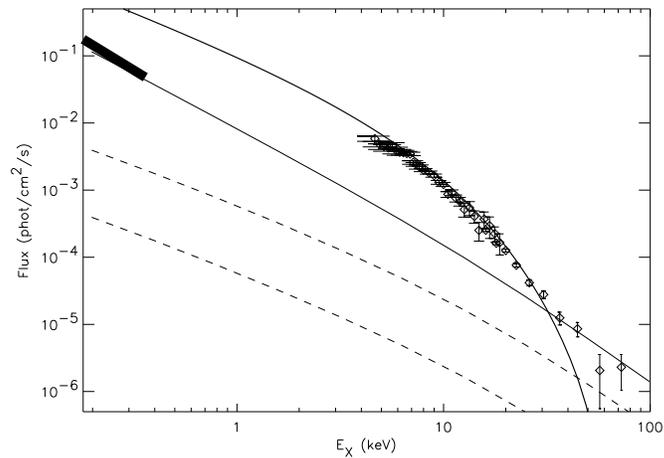,width=9.cm}}
  \caption{\em {Spectrum of ICS radiation from Coma. 
}}
 \end{center}
\end{figure}

With the parameters used to fit the radio and X-ray fluxes, we also 
calculate the gamma ray emission, due to several channels. The results are
plotted in Fig. 3. The thick solid curve is the bremsstrahlung contribution 
of primary electrons, the two solid thin lines represent the fluxes of
gamma rays from pion decay, the dashed lines are ICS fluxes from secondary
electrons and the dashed dotted lines are the bremsstrahlung fluxes 
from secondary electrons. The fluxes produced by secondaries are always 
plotted for $\xi=0.1$ (lower curves) and $\xi=0.01$ (upper curves). 
The EGRET upper limit (Sreekumar et al. 1996 - arrow in the figure) is 
exceeded by a factor 3-4 . 
If a Bohm diffusion coefficient is used, then the maximum energy 
of primary electrons becomes large enough to generate an appreciable flux of 
gamma rays due to ICS. In this case the excess is a factor $\sim 15$.

\begin{figure}[thb]
 \begin{center}
  \mbox{\epsfig{file=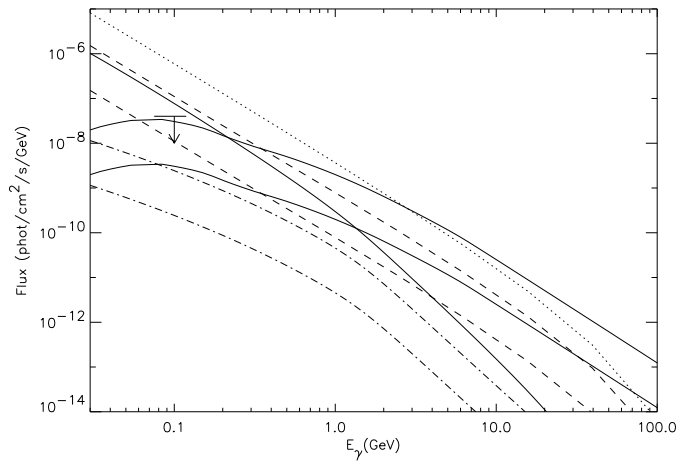,width=9.cm}}
  \caption{\em {Spectrum of gamma radiation from Coma. 
}}
 \end{center}
\end{figure}

Adopting magnetic fields even slightly larger than $0.15\mu G$ causes
the radio radiation to exceed observations if the X-ray flux (e.g. the
cosmic ray normalization) is fixed. Viceversa, if the magnetic field is
increased and we fit the radio observations, the X-ray fluxes become
too small. 

\section{Cluster mergers: alternative views}

In the previous section we emphasized that the {\it conventional
interpretation} of the non-thermal phenomena observed in clusters of
galaxies leads to implications that seem to be in contrast with 
observations: a) small magnetic fields compared with Faraday Rotation
measurements; b) large gamma ray fluxes.

There is an alternative way of explaining observations (Blasi 2000a;
Dogiel 1999) without these problems: the idea consists of the following
key points (Blasi, 2000a): 
1) magnetic fields in the intracluster medium are at $\mu G$ 
level, as indicated by Faraday rotation measurements; 2) radio radiation 
is generated by synchrotron emission of relativistic electrons, with a 
very small energy content; 3) during the merger the electron thermal 
distribution is changed due to resonant interactions with perturbations
in the magnetic field and acquire a non-Maxwellian tail; 4) hard X-rays
are the result of bremsstrahlung of this modified electron distribution. 

Detailed calculations of the development of this tail were carried out 
by Blasi (2000a): all processes responsible for the thermalization,  
resonant interaction with waves, and energy losses were included in the 
form of terms in a time-dependent
non-linear Fokker-Planck (FP) equation, which was then solved numerically. 
This is equivalent to study the process of thermalization of a plasma
in the presence of a perturbed magnetic field. We find that the 
electron distribution is not a Maxwell-Boltzmann distribution, which is a 
rather general result. There are two aspects to keep in mind: first, the
thermalization time of electrons at low energies is extremely short, therefore 
the only region where possible distorsions from a thermal distribution 
can occur is on its tail; second, the FP equation is such that even if the 
energy transfer from the waves to electrons occurs only on the tail 
(lower energy electrons do not resonate), also the rest of the distribution 
is affected. 

As a consequences of the points just stressed, the resonant interaction of
electrons with waves in the intracluster medium mainly results into two 
effects: 1) the bulk of the electrons is heated up (the effective temperature
increases); 2) a non-Maxwellian tail develops.

The results of Blasi (2000a) are plotted in Fig. 4a for the
electron spectra and 4b for the X-ray spectra obtained as bremsstrahlung
emission of the modified electron distribution. The data points are from
BeppoSAX while the upper limits are from OSSE (Rephaeli et al. 1994). 
The different curves refer to different times.

\begin{figure}[thb]
 \begin{center}
  \mbox{\epsfig{file=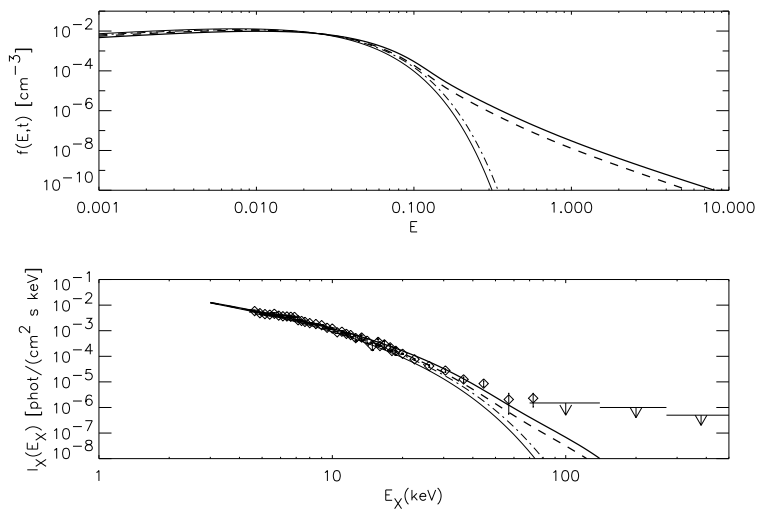,width=10.cm}}
  \caption{\em {{\it a.} Spectrum of electrons. {\it b)} X-ray spectra.
The thin solid lines are for $t=0$, the dashed lines refer to 500 million years
and the thick solid lines refer to 1 billion years. The dash-dotted lines 
refer to the thermal case at the temperature of Coma. The initial temperature
of the gas is 7.5 keV.
}}
 \end{center}
\end{figure}

This alternative model can be tested: since the electron spectrum is changed,
the spectrum of the upscattered photons of the microwave background radiation
(the so-called Sunyaev-Zeldovich effect) is also affected. Blasi, Olinto
and Stebbins (2000) calculated in detail this modified SZ effect and found 
that: {\it i)} accurate measurements of the SZ spectrum should show distorsions
distinguishable from the ordinary SZ effect; {\it ii)} these changes
in the SZ spectra would affect the estimate of the Hubble constant and other
cosmological parameters by $\sim 20\%$; {\it iii)} the masses estimated
from X-rays should be a factor $\sim 2$ smaller than those derived from 
gravitational lensing, sensitive to the total energy (mass plus
thermal energy plus non-thermal energy) in the cluster.

\section{Conclusions}

Radio radiation, hard and soft X-ray excesses and UV radiation have been
observed from several clusters of galaxies, as a clear demonstration of 
the presence of cosmic rays and magnetic fields in the intracluster 
medium. The energetics 
involved, in terms of rate of injection of cosmic rays in clusters, seems 
to be compatible with the ones expected during mergers of clusters of 
galaxies. For this reason, in this paper we investigated the propagation 
and the radiative processes of both primary and secondary electrons and 
of protons, and we calculated the expected fluxes of radio, X and gamma 
radiation, applying our results to the specific case of Coma, for which 
a complete set of multiwavelength observations is available. 

The general conclusion is that primary electrons can provide a good fit
to the observations, provided the merger just ended (or is still ongoing).
The rate of injection of primary electrons needed to fit the X-ray data
is $\sim 2-3\%$ of the total energy available during the merger, and 
a magnetic field of $0.15\mu G$ is required to fit the radio data. The
flux of UV radiation is automatically obtained in the right order of 
magnitude. All these fluxes are rapidly fading away with time, due to energy
losses of high energy electrons. The contribution of the secondary electrons
is slightly smaller but time independent, due to the confinement of cosmic
ray protons on cosmological time scales. 

The main consequence of the model is that the gamma ray fluxes above $\sim 100$
MeV exceed the EGRET upper limit by a factor between 3-4 to $\sim 15$
depending on the diffusion coefficient, which determines the maximum energy 
of the primary electrons. Besides this, the use of a magnetic field of
$\sim 0.15 \mu G$ seems to be inconsistent with the results of recent
Faraday rotation measurements (Eilek 1999; Clarke et al. 1999). 

There are two things to do in order to clarify the situation: 1) carry out
a new measurement of the gamma ray fluxes in the GeV range, in order to
check that EGRET did not just miss the detection; 2) carry out gamma ray
measurements in the range $>100$ GeV, as proposed by Blasi (1999). These
measurements can give precious information about the non-thermal content 
of clusters and they can be carried out even with current detectors like
STACEE (Ong 1998).

A more theoretical approach consists in looking for alternative interpretations
of the non-thermal radiation. We discussed this possibility in section 3,
where we illustrated the calculations of Blasi (2000a), relative to 
the spectrum of a population of electrons that  
thermalize in the presence of a perturbed magnetic field. By solving the 
FP equation, Blasi (2000a) found that the electron spectrum presents an high 
energy tail, and that the bremsstrahlung emission from these
electrons can explain the whole spectrum of X-rays, from the thermal region
to the hard X-ray region. Magnetic fields compatible with the ones
obtained by Faraday rotation measurements and low gamma ray fluxes can be 
accomodated naturally within this approach.

Moreover it is possible to prove (or disprove) this alternative model by
looking at precision measurements of the Sunyaev-Zeldovich (SZ) effect: the 
electron distribution, modified by the resonant interactions with waves
in the intracluster magnetic field produces a modified SZ effect that can be
distinguished by the ordinary one, as discussed by Blasi, Olinto and Stebbins
(2000).

{\bf Aknowledgments} This work was supported by the DOE and the 
NASA grant NAG 5-7092 at Fermilab.

\section{References}

\reff
Berezinsky, V.S., Blasi, P., Ptuskin, V.S.: 1997 Astrophys. J. {\bf 487}, p.
 529.

\reff 
Blasi, P.: 2000 submitted to Astropart. Phys.

\reff
Blasi, P.: 2000a Astrophys. J. Lett. {\bf 532}, p. L9. 

\reff
Blasi, P.: 1999 Astrophys. J. {\bf 525}, p. 603.

\reff
Blasi, P., Colafrancesco, S.: 1999 Astropart. Phys. {\bf 12}, p. 169.

\reff
Blasi, P., Olinto, A.V., Stebbins, A.: 2000 Astrophys. J. Lett. in press.



\reff
Clarke, T.E., Kronberg, P.P., B\"{o}ringer, H.: 1999 
in Ringberg Workshop on ``Diffuse Thermal and 
Relativistic Plasma in Galaxy Clusters'', Eds: H. B\"{o}ringer, L. Feretti,
P. Schuecker, MPE Report No. 271.

\reff
Colafrancesco, S., Blasi, P.: 1998 Astropart. Phys. {\bf 9}, p. 227.

\reff
Deiss, R.M., Reich, W., Lesch, H., Wielebinski, R.: 1997
A\&A {\bf 321}, p. 55.

\reff
Dogiel, V.A.: 1999 in Ringberg Workshop on ``Diffuse Thermal and 
Relativistic Plasma in Galaxy Clusters'', Eds: H. B\"{o}ringer, L. Feretti,
P. Schuecker, MPE Report No. 271, 1999.

\reff
Eilek, J.A.: 1999 preprint astro-ph/9906485,  to appear in 
'Diffuse Thermal and Relativistic Plasma in Galaxy Clusters', 1999, 
Ringberg Workshop, Germany, MPE-Report.

\reff
Ensslin, T.A.: 1999 Proceedings of {\it The Universe at Low Radio 
Frequencies}, ASP Conference Series, (preprint astro-ph/0001433).



\reff
Ensslin, T.A., Lieu, R., Biermann, P.: 1999 A\&A {\bf 344}, p. 409.

\reff
Feretti, L., Dallacasa, D., Giovannini, G., Taglianai, A.: 1995 A\&A {\bf 302},
p. 680.

\reff
Fusco-Femiano, R., Dal Fiume, D., Feretti, L., Giovannini, G., Grandi, P., 
Matt, G., Molendi, S., Santangelo, A.: 1999 Astrophys. J. Lett. {\bf 513}, 
p. L21.

\reff
Levinson, A.: 1994 Astrophys. J. {\bf 426}, p. 327.

\reff
Lieu, R., Ip, W.-H., Axford, W.I., Bonamente, M.: 1999 Astrophys. J. 
{\bf 510}, p. 25.

\reff
McClements, K.G., Dendy, R.O., Bingham, R., Kirk, J.G., Drury, L. O'C.: 1997
MNRAS {\bf 291}, p. 241.

\reff
Rephaeli, Y., Ulmer, M., Gruber, D.: 1994 Astrophys. J. {\bf 429}, p. 554.






\reff
Sarazin, C.L.: 1999 Astrophys. J. {\bf 520}, p. 529.


\reff
Ong, R.A.: 1998 Phys. Rep. {\bf 305}, p. 93.

\reff
Sreekumar, P. et al.: 1996 Astrophys. J. {/bf 464}, p. 628.

\reff
Takizawa, M., Naito, T.: 2000 preprint astro-ph/0001046 (accepted for 
publication in Astrophys. J.).

\reff
Volk, H.J., Aharonian, F.A., Breitschwerdt, D.: 1996 Space Sci. Rev. {\bf 75},
p 279.

\reff

\end{document}